%% file: gpu-rtcg.tex
\documentclass{elsarticle}

\usepackage[utf8]{inputenc}
\usepackage{graphicx}
\usepackage{color}
\usepackage{verbatim}
\usepackage{url}
\usepackage[numbers]{natbib}
\usepackage{hypernat}
\usepackage{listings}
\usepackage{multirow}

\usepackage{tikz}
\usetikzlibrary{positioning}
\usetikzlibrary{calc}
\usetikzlibrary{chains}
\usetikzlibrary{shapes.misc}
\usetikzlibrary{shapes.symbols}
\usetikzlibrary{shapes.geometric}
\usetikzlibrary{fit}
\usetikzlibrary{shadows}

\pgfdeclarelayer{background}
\pgfdeclarelayer{foreground}
\pgfsetlayers{background,main,foreground}

\colorlet{codeback}{gray!20}

\usepackage[margin=10pt,font=small,labelfont=bf,labelsep=period]{caption}

\usepackage[letterpaper,margin=1in,includefoot]{geometry}

\usepackage{hyperref}

\begin{document}


\begin{frontmatter}

\title{PyCUDA and PyOpenCL: A Scripting-Based Approach to GPU~Run-Time~Code~Generation}

\author[cims]{Andreas Kl\"ockner}
\ead{kloeckner@cims.nyu.edu}
\author[mit,rowland]{Nicolas Pinto}
\ead{pinto@mit.edu}
\author[eecsberk]{Yunsup Lee}
\ead{yunsup@eecs.berkeley.edu}
\author[eecsberk]{Bryan Catanzaro}
\ead{catanzar@eecs.berkeley.edu}
\author[vs,redwood]{Paul Ivanov}
\ead{pi@berkeley.edu}
\author[eeceohio]{Ahmed Fasih}
\ead{fasih.1@osu.edu}

\address[cims]{Courant Institute of Mathematical Sciences, New York University, New York, NY 10012}
\address[mit]{McGovern Institute for Brain Research, Massachusetts Institute of Technology, Cambridge, MA 02139}
\address[rowland]{Rowland Institute, Harvard University, Cambridge, MA 02142}
\address[eecsberk]{Electrical Engineering and Computer Sciences, University of California, Berkeley, CA 94720}
\address[vs]{Vision Science Graduate Program, University of California, Berkeley, CA 94720}
\address[redwood]{Redwood Center for Theoretical Neuroscience, University of California, Berkeley, CA 94720}
\address[eeceohio]{Department of Electrical and Computer Engineering, The Ohio State University, Columbus, OH 43210}

\begin{abstract}

High-performance computing has recently seen a surge of interest in
heterogeneous systems, with an emphasis on modern Graphics Processing Units
(GPUs). These devices offer tremendous potential for performance and efficiency
in important large-scale applications of computational science. However,
exploiting this potential can be challenging, as one must adapt to the
specialized and rapidly evolving computing environment currently exhibited by
GPUs. One way of addressing this challenge is to embrace better techniques and
develop tools tailored to their needs. This article presents one simple
technique, \emph{GPU run-time code generation} (RTCG), along with PyCUDA and
PyOpenCL, two open-source toolkits that support this technique.

In introducing PyCUDA and PyOpenCL, this article proposes the combination of a
dynamic, high-level scripting language with the massive performance of a GPU as
a compelling two-tiered computing platform, potentially offering significant
performance and productivity advantages over conventional single-tier, static
systems. The concept of RTCG is simple and easily implemented using existing,
robust infrastructure. Nonetheless it is powerful enough to support (and
encourage) the creation of custom application-specific tools by its users. The
premise of the paper is illustrated by a wide range of examples where the
technique has been applied with considerable success.

\end{abstract}


\end{frontmatter}



\section{Introduction}
\label{sec:intro}

Graphics Processing Units (GPUs)
\cite{dally_merrimac_2003,lindholm_nvidia_2008,seiler_2008} promise tremendous
advantages in throughput over conventional processor architectures, ideally
resulting in a large reduction of execution time for suitable compute-~or
bandwidth-bound algorithms. However, execution time is not the only time scale
to consider when comparing computer architectures. Indeed, the development time
for a scientific code will, in many cases, be a significant fraction of its
useful lifespan. GPUs now threaten to tip this balance even further out of the
programmer's favor, through the following four factors.

First, there is still much change going on in the area of massively parallel
processors. These changes are driven by many factors--chip manufacturing processes
change, new ideas and abstractions in hardware and software emerge and
disappear at a rapid pace, market conditions change. Programs that work
well on last year's machines may not continue to represent optimal choices
today. While the OpenCL standard
\cite{opencl_2008} provides a unified abstraction for conceptualizing
widely divergent hardware architectures, performance portability
remains a difficult problem, both between contemporary competing
architectures available today, as well as between today's
architectures and those of the future.  Even though some patterns are emerging, the
world is still very far from having settled on a programming model for
massively parallel machines--a model that is as stable as the one we have
enjoyed on CPUs for the last few decades.

Second, GPU code is very sensitive to seemingly innocent changes.  Hardware
implementation details are much more visible and have a much greater
performance effect in GPU programs than they do in
today's CPU programs. Relative component clock rates, bus widths, vector
widths, memory and buffer sizes all have an immediate impact on a successful
code. The \emph{very premise} of GPU computing is to try and find a better use
for the silicon tied up in the caching, speculation and out-of-order execution
that frees a modern CPU developer from having to worry about hardware
peculiarities.  We therefore expect that GPU developers will continue to be
exposed to these details.

Third, and potentially a corollary of the last point, GPUs offer many more
implementation choices, and often little guidance on which choice may lead to
efficient code. It is not uncommon to see differences of an order of
magnitude in execution time between codes that accomplish the same basic task.
This is not likely to occur on a current-generation CPU, where, with few
exceptions, ``reasonably coded'' and ``highly optimized'' fall within at most a
factor of two or three of each other.

The fourth and possibly worst factor is that GPU development tools are in their infancy. Many
years have been spent creating development tools that help the CPU developer
achieve high productivity. These tools range from high-level languages and
libraries that allow the programmer to deal in convenient abstractions, to
optimizing compilers, debuggers, and profilers, which likewise shield the
programmer from having to deal with the full complexity of the hardware. Many
of these tools are either unavailable, inadequate or rudimentary on today's
parallel architectures.

We propose that \emph{GPU run-time code generation} (``RTCG'') helps the
programmer reclaim a significant share of the productivity lost to these
factors. By GPU RTCG, we mean the ability to seamlessly execute arbitrary,
generated low-level C (or C-like) source code for high-volume computational
tasks in the context of the generating program. In the form described in this
paper, the generation and execution of the low-level code is performed from a
high-level scripting language.  By the term ``scripting language'' or
``high-level language'', we mean a language that
\begin{itemize}
  \item enables various programming paradigms (e.g. functional, procedural,
    object, aspect, etc.),
  \item is dynamically typed,
  \item includes error reporting facilities,
  \item manages resources automatically,
  \item offers comprehensive built-in functionality,
  \item requires no user-visible compilation (i.e. suitable for interactive use), 
    and
  \item works well as a ``glue language'' for lower level building blocks.
\end{itemize}
The family of major general-purpose scripting languages at the time of this 
writing includes Python \citep{vanrossum_python_1994}, Ruby \cite{flanagan_ruby_2008},
Lua \citep{ierusalimschy_programming_2006}, JavaScript \citep{eich_javascript_2005}
and numerous others.

The present work describes lessons learned from many earlier approaches. GPU
RTCG is a form of ``metaprogramming'': instead of directing computer code
immediately at a problem, one directs code at the creation of and reasoning
about 
other codes which then solve the problem at hand. It is not
initially clear that this additional level actually results in any tangible
gain, but we defer this discussion to the later parts of this article.
For now, it should suffice to say that we are by no means the first to apply
the basic principle. Today, perhaps the most common mechanism used to implement
metaprogramming ideas is the template mechanism of the C++ programming
language. Many things have been implemented in this effective (if cumbersome)
way: Expression evaluators \cite{veldhuizen_1997}, parser generators
\cite{guzman_spirit_2008}, even entire PDE solver frameworks
\cite{reynders_pooma_1996,prudhomme_domain_2006}. The template-based technique
is, however, constrained to being applied at the time when the software is built,
which limits its usefulness. A variety of ways have been devised to circumvent
this restriction, reaching from assembly of small prefabricated pieces into a
full code \cite{frigo_2005}, to build-time evaluation of different code
versions \cite{whaley_automated_2001}.  It should further not be forgotten that
the Lisp programming language already brought the fundamental insight of the
von~Neumann architecture, namely that `code is data', to higher-level languages
in the early 1960s \cite{mccarthy_lisp_1962}, albeit not necessarily with
computational efficiency as the primary target.

In the context of GPUs, metaprogramming has so far been applied mainly in a
graphics and image processing context \cite{lejdfors_implementing_2006,
wedekind_machine_2008} and to ease the use of a standard rendering pipeline for
general-purpose uses \cite{tarditi_accelerator_2006}.  Other projects focus on
generating GPU code using a compile-time C++-based framework
\cite{mccool_metaprogramming_2004,mccool_data-parallel_2006}.

\begin{figure}
  \centering
  \begin{tikzpicture}[
    font=\sffamily\small,
    actor/.style={cylinder,
      fill=none,draw=green,thick,
      cylinder uses custom fill=true,
      cylinder end fill=green,
      cylinder body fill=green!50,
      shape aspect=.5,
      text height=1em,
      text depth=0.5ex,
      text centered,
      },
    object/.style={
      rectangle,
      minimum height=3.5ex,
      inner xsep=3mm,
      fill=lime!50,
      draw=lime,
      text height=1em,
      text depth=0.5ex,
      thick,
      },
    hlbox/.style={
        rectangle,fill=#1!30,draw=#1,opacity=0.5,thick
    },
    button/.style={
      rounded rectangle,
      bottom color=gray!60,
      top color=gray!20,
      inner sep=2mm,
      draw=gray!30,very thick,
      },
    ]
    \node [object] (idea) at (0,0) {Idea} ;
    \node [actor,right=0.3 of idea] (py) {Scripting Code} ;
    \draw [->,thick] (idea) -- (py) ;

    \node [object] at (-4,-1.8) (cu) {GPU Code} ;
    \draw [->,thick] (py.east) -| ++(0.75,-0.9)  -| ++(-10.5,0) |- (cu.west) ;
    \node [actor,right=0.3 of cu] (nvcc) {GPU Compiler} ;
    \draw [->,thick] (cu) -- (nvcc) ;
    \node [object,right=0.3 of nvcc] (cubin) {GPU Binary} ;
    \draw [->,thick] (nvcc) -- (cubin) ;
    \node [actor,right=0.3 of cubin] (gpu) {GPU} ;
    \draw [->,thick] (cubin) -- (gpu) ;
    \node [object,right=0.3 of gpu] (result) {Result} ;
    \draw [->,thick] (gpu) -- (result) ;

    \begin{pgfonlayer}{background}
      \node [hlbox=green,fit=(idea) (py),inner sep=2.5mm] (human) {};
      \node [left=0.3 of human,button] {Human} ;
    \end{pgfonlayer}

    \begin{pgfonlayer}{background}
      \node [hlbox=red,fit=(cu) (result),inner sep=2.5mm] (machine) {};
      \node [right=0.3 of machine,button] {Machine} ;
    \end{pgfonlayer}

  \end{tikzpicture}
  \label{fig:gpu-code-generation}
  \caption{Operating principle of GPU code generation.}
\end{figure}
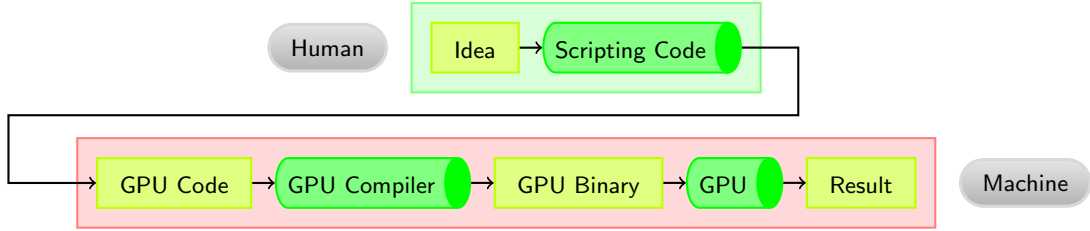

Further, this work can be seen in the context of recent efforts
\cite{lengauer_2004} to promote program generation as a mainstream idea. In
comparison however, we are choosing a decidedly simple approach that values
pragmatism over theoretical appeal: Why should we invent new tools from scratch
when good results are achievable using a scripting language with a GPU and a C
compiler?  Curiously, many previous authors give up the immeasurable advantage
of being able to generate code \emph{at run time} all too easily.  This
capability is the main point of this article.

The text is organized as follows: We begin by giving a very brief overview of
how GPUs differ from other computing platforms, first from the point of view of
hardware in Section~\ref{sec:gpu-hardware}, then from that of software in
Section~\ref{sec:gpu-software}. We continue in
Section~\ref{sec:problem-overview} by providing a sampling of problems arising
from a GPU's special structure where GPU RTCG can be profitably applied.
Section~\ref{sec:approach} then describes a scripting-based approach to these
problems that is supported by our open-source \emph{PyCUDA} and \emph{PyOpenCL}
toolkit.  Section~\ref{sec:applications} describes how a number of applications
from varied disciplines have benefited from our approach. Finally, in
Section~\ref{sec:conclusions}, we close with a few remarks and ideas for future
work.



\section{GPU Hardware: A Brief Introduction}
\label{sec:gpu-hardware}

In the early days of GPU programming, the programmer had to repurpose
marginally programmable fixed-function graphics hardware for computing purposes
by a variety of methods \cite{owens_survey_2007}.  With today's generation of
GPUs, this is not true any more. Instead, GPUs should be viewed as
general-purpose highly parallel processors that are designed for a different
type of target workload than current CPUs, and ``GPU'' becomes just a
convenient moniker for this type of technology. For CPUs, the set of design
workloads typically includes web browsers, word processors and a diverse
collection of other desktop programs--characterized by high complexity and
marginal potential for parallelization. GPUs, on the other hand, are aimed at
applying uniform, moderately complex operations to large volumes of data.  This
constitutes a special (but important) subset of parallel computation, often
called data-parallel or ``stream'' computing \cite{gpu_stream_computer_2003}.

One of the most significant problems that modern processor design needs to
address is the slowness of memory. While there have been significant advances
in latency and access speed to affordable, large-scale, off-chip random access
memory, these advances have in no way kept pace with the progress made in the
throughput of processor cores. Variants of Moore's Law predicted this latter
progress to be exponential in nature, and so far reality has kept pace with
prediction.  Consequently, we now deal with a growing gap between the
computational capabilities of parallel processors, and the speed at
which they can access the data required to perform computation. The time between
the issuing of a memory request by a core and the subsequent response from
off-chip memory can be very long, up to hundreds or even thousands of
processor cycles, and the gap is widening.

While bandwidth can be increased to some extent by widening and improving the
memory interface, latency cannot, as it is a fundamental property of the type
of memory. Obviously, the design workloads for CPUs are very vulnerable to
memory delays, and therefore CPU designers tend to take extreme measures to
mitigate their effects. Three types of strategies are particularly popular
here: First, include large amounts of fast cache memory on the chip to avoid
having to wait for off-chip memory at all. Second, engage in many forms of
prediction and speculation to make sure that required data is already present
on-chip when it is needed. And finally, reorder the instruction stream to
lessen the impact of memory-related stalls.

It is apparent that the hardware implementation of all these strategies can
easily occupy large amounts of silicon. In contrast, the target workloads for a
GPU are much less vulnerable to memory-related stalls. Since GPUs aim to apply
similar operations to large amounts of data, exact ordering is less important.
This allows the use of a much larger number of execution contexts, each of
which may occupy a functional (i.e. floating-point or integer) unit whenever it
has data available. While the management of large numbers of contexts is
nontrivial in itself, the associated management logic is less expensive to
implement than the CPU's strategies, freeing a GPU to dedicate much more chip
space to functional units, further increasing parallelism.

This abundance of functional units confronts GPU designers with yet another
interesting challenge. Context management logic grows strongly superlinearly
with the number of contexts it manages. One set of central logic that would
manage the execution of all contexts on all functional units on the chip would
be prohibitively large.  This, together with physical limits of on-chip signal
propagation speed, strongly suggests dividing up the available chip are into
individual sub-processors, each of which manages a more limited set of
execution contexts. It is the same thinking that drives heavyweight CPUs
towards integrating multiple cores on a single die. Likewise, modern GPUs
contain tens of management subdomains, each of which may manage hundreds of
execution contexts. (These subdomains are called `compute units' by OpenCL,
`multiprocessors' by Nvidia, and simply `cores' by others. Execution contexts
are called `threads' by Nvidia and `work items' by OpenCL.) To further improve
the functional-unit-to-control-logic ratio and reach the cited width of
hundreds of contexts per subdomain, most GPUs are built as relatively
wide SIMD (Single Instruction Multiple Data) vector machines.

The chip$\rightarrow$unit$\rightarrow$context hierarchy has a twofold effect on
GPU software: First, each unit is typically designed to operate independently
of its siblings, limiting communication to contexts executing on the same unit.
Second, programs must explicitly specify how to use each level of parallelism,
typically by providing a suitable decomposition of an index space.  Together
with the remaining possibility of sequential execution, this poses the problem
of \emph{loop slicing}. Given a sequential description of the algorithm as a
set of nested loops, loop slicing refers to the combined process of
\begin{itemize}
\item identifying loop axes that can serve as parallelization indices,
\item assigning loop axes to available parallelization axes, such as
  compute units, execution context numbers within a unit, and SIMD lanes,
\item interchanging loop orders to achieve a more beneficial order
  of memory accesses, and lastly,
\item finding size restrictions on each loop axis, and splitting
  axes as necessary.
\end{itemize}
Observe that each of the above steps may depend on the outcome of all the
others, resulting in a complicated joint optimization problem.  The purpose of
the remainder of this article is to explore these (and other) software
challenges and propose solutions for some of them.



\section{GPU Software Creation}
\label{sec:gpu-software}

Writing efficient software for GPUs using low-level programming
environments such as OpenCL and CUDA requires mapping computation
directly onto GPU architectural structures.  This mapping depends
critically on a host of architectural parameters, such as:
\begin{itemize}
  \item the width and number of available compute units,
  \item the amount of available register file state on chip
  \item the amount of available on-chip buffer memory,
  \item the speed of various access patterns to on- and off-chip memory,
  \item the ratio of available memory bandwidth to compute bandwidth,
  \item the latency and bandwidth between the host (CPU) and the device (GPU), and
  \item the instruction scheduling details of the processor in use.
\end{itemize}

There are many possible mappings of a given computation onto GPU hardware
constructs, each with unique performance characteristics.  As might be
expected, the mapping process is complicated, often involving trade-offs
between efficiency in one dimension versus another. In many cases, the
programmer making these trade-offs has incomplete information on the factors
involved. For example, design details of the compute device may be unavailable
to the programmer.  But even if they are, program execution in massively
parallel processors is a complicated and non-local process that may defy easy
comprehension even by the processor's designers.

GPU programming therefore relies extensively on experimentation and
microbenchmarking to overcome missing knowledge of causes by obtaining
measurements of symptoms.  As a software developer, this is a very
unsatisfying place to be in: the obtained results may not be robust to
changes of hardware, problem sizes or other parameters.  The
optimization problem is further complicated by the fact that GPUs by
their very nature consist of simple cores that omit features like
speculative execution and out-of-order execution which might help make
some decisions less critical.  Further, this experimentation and
benchmarking is generally tedious work that needs to be carried out
systematically, consistently and repeatably. It is therefore not
far-fetched to wish for these tasks to be automated. From there, it is
a small step to \emph{metaprogramming}, the automated reasoning about
programs, and RTCG.



\section{Problems Solved by GPU Run-Time Code Generation}
\label{sec:problem-overview}

This section is devoted to describing a number of issues that are commonly
faced when programming a GPU. In each case, we point out how a GPU RTCG
strategy can be used to address these issues in a natural and straightforward
manner.

\subsection{Automated Tuning}

During the creation of a GPU program, it is natural for the programmer to come
up with a number of variants of a given code, each of which will be observed to
have certain properties regarding data layout and computation speed. The
conventional approach to code tuning then calls for the fastest variant to
survive, while the others will be discarded. This is not necessarily a
desirable course of action, as information is lost. Instead, it seems more
appropriate to retain as many of these variants as is practical, assuming that
they hold at least some promise. Further, each variant may have a number of
tunable parameters, such as loop lengths, block sizes, etc. Retaining variant
information permits choosing the best one from a reasonable-size pool of
candidates in an automated fashion, guided by some metric such as execution
speed. This is the basic premise of automated tuning, which is trivially
enabled by GPU RTCG. Further, automated tuning is not just enabled by RTCG, it
is enabled \emph{at the right time}--namely at run time--when complete
information is available.  We present three examples illustrating the type of
choices optimally resolved by automatic tuning.

The first and perhaps the most important choice in GPU algorithm design is that
of loop slicing, as explained in Section~\ref{sec:gpu-hardware}. Even loops
that are trivially linear on the CPU must typically be subdivided into several
levels for the GPU to be efficient, with levels corresponding to SIMD lanes,
execution units, as well as serial execution.  For some algorithms such as
matrix multiplication, loop slicing is important even on the CPU to preserve
locality of access and thereby the efficiency of on-chip caches. Since GPUs
have even less cache and even more slicing levels, getting the loop slicing
right is of paramount importance to obtaining reasonable performance.

Second, today's GPU architectures have \emph{user-managed on-chip memories}. Upon
creation of a code, it is often not obvious which pieces of data will yield the
most benefit from low latency local storage. It is almost certain that on-chip
memory will remain a scarce resource for the foreseeable future. Thus, peak
performance necessitates trade-offs that adapt to the hardware situation at
hand.

Third, GPU architectures are built to utilize large amounts of DRAM
bandwidth.  Contention for DRAM bandwidth is therefore a critical
performance limiting factor.  Due to the characteristics of DRAM and
memory controller architectures, efficiently utilizing DRAM bandwidth
requires paying close attention to how memory is accessed.
Consequently, GPU performance tuning often centers around optimizing
memory access patterns, which entails changing data structure layouts
as well as the mapping of computation onto GPU threads. This
optimization problem involves making trade-offs between how
computation is performed and how data is stored.  In the absence of
tools for examining these trade-offs, the programmer must manually
restructure their code and data to explore the tuning space, which is
tedious and error prone.

\subsection{The Cost of Flexibility}

Flexibility is commonly seen as a desirable feature of a computer code--where
``code'' usually means a user-facing executable. The more functions a certain
executable can perform without having to be modified, the better. Yet there
exists a flexibility versus performance trade off. As an example that is the
polar opposite of flexibility, one may consider an optimized code that can only
multiply matrices of a certain size. No matter how fast or otherwise attractive
such a code may be, unless the user's desired application requires matrix
multiplications of this size, it is entirely useless. Thus almost all computer
codes are built with at least some flexibility.

It should then be realized that flexibility comes at a cost: Constants
get replaced by variables, formerly fixed loop trip counts become variable, and
quite generally a compiler has less knowledge available at compile time, making
its optimizer less effective. The process of removing such flexibility, on the
other hand, is generally frowned upon and derisively called ``hardcoding''.
We feel, however, that this point of view has no merit once run-time code generation is
available, as one is at liberty to generate code for exactly one purpose--any
extra flexibility is likely just unneeded ballast.

In \emph{compile-time} metaprogramming frameworks, hardcoding is sometimes
replaced by generating a large number of potentially needed code variants ahead
of time by considering anticipated needs for different problem sizes, data types,
etc.  Once the number of variants surpasses ``a few'', the costs of this
approach quickly become very significant both in compilation time and memory
footprint of the executable. In comparison, GPU RTCG suffers no such scaling
penalty: It can use information available only at run time to cut down the
number of variants that need to be generated, it can use caching to amortize
the cost of finding the optimal code, and unused code variants can be disposed
of immediately.

\subsection{High-Performance Abstractions}

Nearly all computer programs are built in `layers', where each individual layer
solves a certain subproblem and presents a more abstract, `higher-level'
interface to surrounding layers. This is good engineering practice, as it
allows for the partitioning of a big problem into many smaller ones, and it enables reuse
of engineering effort. In some cases, this layering is easily achieved and
results in very little loss for the `consumer' of the interface.  In other
cases, such abstractions can be made uneconomical by coding circumstance. We
will first look at examples of how this might happen, and then at what RTCG
does to improve the situation. One common instance of uneconomical abstractions
occurs when a consumer of an interface needs to specify details about an
operation that is to be performed on large volumes of data, as part of an inner
loop in the abstraction. As a trivial example, consider an abstract form of
vector addition allowing a variety of scalar types.

One simple run-time technique is the use of function pointers (or
equivalently, virtual methods). In the frame of our example, the cost of a
scalar addition is far smaller than that of a call through a function pointer,
often by several orders of magnitude.  The reason for this is that such a call
may may defeat prediction logic and stall the execution pipeline. The use of a
computed call for a single additions is therefore impractical, and one would
need to amortize its cost over many operations (scalar additions in the
example), if this is feasible. In addition to the added complexity, this
approach may not be viable on certain GPU architectures that are still quite
common as of this writing. (OpenCL 1.1 for example specifically excludes
function pointers.)

The disadvantages of the function pointer approach drove the development of
mechanisms for compile-time-polymorphism on the CPU and the GPU. In C++, this
is achieved through the use of class and function \texttt{template}s. If the
user's customization is assumed to be known at compile time, the compiler can
make use of that knowledge and generate efficient code. In our example, the
vector addition would be written with respect to an unspecified type, relying
(for example) on the assumption that the underlying scalar supplies addition.
The type of the scalar is required to be known at compile time, and hence the
compiler can statically find the addition routine and substitute (``inline'')
its use, ideally eliminating all overhead. This is a popular approach, but it
has two shortcomings: First, it requires early concretization.  In the example,
all desired uses of the vector addition code have to be known before the
program is run. Second, the C++ \texttt{template} mechanism in particular
responds unfavorably to complexity growth. It makes simple things like type
substitution quite easy.  But \texttt{template}s alone, even without the rest
of C++, form a fully capable--if awkward--programming language
\cite{veldhuizen_templates_2003}, and some implementers have seen this as an
invitation to do rather advanced things with them. While such use validates the
need for a meta-level where code is able to reason about other code, the actual
end results in this case tend to be both brittle and complicated.

The ideal solution would be a compromise of these two. Function pointers are
simple, flexible and do not require early concretization, while templates have
very little overhead. By removing the distinction between `compile time' and
`run time', RTCG fills this void. Once RTCG is available, appropriate code can
be generated whenever a different requirement arises, leading to flexibility.
RTCG code is also fast--it can do away with any sort of flexibility, because it
can safely be considered ``single-purpose''. Further, code generation can be
seen as a text processing task. Since one is not limited in the choice of tools
with which to perform this generation, RTCG-based codes can be as simple as
possible and respond favorably to complexity growth.

\subsection{GPUs and the Need for Flexibility}

As a final comment, it should be emphasized that in the past, due to the
associated development complexity especially for C++\,-based techniques,
metaprogramming was restricted to high-need applications.  The cost of
metaprogramming outweighed the disadvantages of ``hardcoding'' only for the
largest of projects.

GPUs however democratize this need, as they put a larger penalty on inflexible,
untuned code. By deciding to perform a GPU port of an algorithm, one implicitly
states that one is willing to trade some implementation effort for a
substantial performance gain. As explained above, finding a good implementation
is often nontrivial, and therefore the potential gain from RTCG is large. In
other words, GPUs increase the relative cost of not using metaprogramming
techniques, and therefore it is likely that code generation and techniques like
it will see much wider adoption. However, good tools are required to allow the
broadest possible cross-section of developers to take advantage of RTCG.



\section{PyCUDA and PyOpenCL: A Scripting-Based Approach to GPU RTCG}
\label{sec:approach}

We have seen in the previous section that GPU RTCG solves a number of pressing
problems in the development of high-performance compute-oriented codes.  In
this section, we present the basic functional principles of our practical and
mature open-source toolkits for supporting GPU RTCG.

As already suggested by their naming, PyCUDA connects the high-level Python
programming language \cite{vanrossum_python_1994} with the Nvidia CUDA compute
abstraction \cite{cuda_prog_2008}, and PyOpenCL connects Python with the
OpenCL \cite{opencl_2008} industry standard compute abstraction. At least the
choice of Python deserves brief justification at this point. The dominating
factor in choosing a high-level, dynamic programming language over a
potentially better-performing, low-level, static one is the complementarity of
tasks between the GPU and the host processor. The GPU is optimally suited to
carrying out throughput-oriented parts of a program, namely the part that would
have conventionally constituted the `inner loops'. Freed from this duty, the
CPU now is responsible for ``only'' control and communication (including, e.g.,
disk input/output). In other words, it now works at a higher level of
abstraction. Therefore a high-level scripting
language (such as Python) can perform this higher-level job equally well or
better, simply because the performance demands are reduced, and both code
generation and execution control can be of considerable complexity. Control
input is needed by the GPU about once every millisecond, and code generation is
needed far less frequently. A Python-based GPU compute code will have no
trouble realizing the same full performance potential of GPU hardware as a
C-controlled GPU compute code, but with much less effort on the part of the
programmer. This reduction in effort is achieved in many ways--for example,
data types and resources are managed by the language itself instead of by a
human, also closures and other high-level constructs are available.  Relatedly
we would like to emphasize that PyCUDA and PyOpenCL do not inhabit Python's software
ecosystem by themselves: a large number of packages for such diverse purposes as
plotting, computer algebra, or optimization are available easily and under
liberal licenses \cite{langtangen_python_2009}. Significantly, the combination
of our software with the
\texttt{mpi4py} package \cite{dalcin_mpi_2005}
allows a straightforward use of hybrid shared-memory GPU-based and
distributed-memory MPI-based parallelism. The easy availability of a multitude
of packages contributes to making scripting languages more productive than
their conventional compiled counterparts. Scripting languages such as Python or
even MATLAB are already popular for exploratory prototyping, but in combination
with a GPU, their usefulness extends well into the territory of `full-scale'
production codes.

Our packages themselves are built from multiple levels. At the lowest level,
each makes the {\em entirety} of the underlying run-time system available from
Python by introducing a thin object-oriented shell. In this context, we would
like to emphasize the word ``entirety'': every feature of the CUDA and OpenCL
run-time systems is accessible from Python, including e.g. textures/images,
OpenGL interaction, zero-copy host memory mapping, timing, and control of
host/compute device parallelism.

While this low-level interface translation is relatively straightforward, care
was taken to make the interface a ``good citizen'' of the high-level-language
system: Memory allocation and resource management concerns are handled
automatically in close coordination with the Python garbage collector, avoiding
spurious resource shortages. Entities such as textures, code modules, and
compute devices are reflected into Python using object-oriented terms,
providing better abstraction than the low-level C interface. Errors are
detected and reported automatically. Further, programmers of high-level
languages expect that their programs do not abort upon executing erroneous
code, that most error conditions are recoverable and that useful feedback is
available on what happened that caused the error.  Our packages satisfy these
expectations. Care is taken however that these automatisms do not turn into a
liability.  For example, a program under tight memory constraints may not have
the luxury of allowing automatic resource management. For this use case, we
still allow the user to manually control deallocation of resources.
Further, memory allocation on the device understands that the
host-level garbage collector might be able to help satisfy a memory
request by freeing up otherwise held device memory.

\begin{figure}[ht]
  \centering
  \begin{tikzpicture}[
    tbox/.style={minimum size=6mm,text height=1.5ex,text depth=0.25ex},
    stepbox/.style={rounded rectangle,thick,tbox,minimum width=4.5cm},
    sidestepbox/.style={rounded rectangle,thick,tbox,minimum width=3.5cm},
    objbox/.style={rectangle,thick,tbox},
    myarr/.style={thick},
    font=\sffamily\small,scale=0.7,
  ]
    \node[stepbox,draw=red,fill=red!40]
      (edit) {Edit} ;

    \draw [fill=gray!20,draw=none] (edit.east) +(1.25,0.9) rectangle +(10.5,-4.8)
      node [pos=0.95,anchor=south east] {PyCUDA};

    \node[stepbox,draw=green,fill=green!40,below=0.5 of edit]
      (run) {Run} ;

    \draw [->,myarr] (edit) -- (run) ;

    \node[objbox,draw=orange,fill=yellow!40,below=0.5 of run,minimum width=3.8cm]
      (smod) {GPU Source Module} ;
    \draw [->,myarr] (run) -- (smod) ;

    \draw [->,myarr] (smod.east) -- +(0.5,0) ;

    \node[diamond,aspect=4,draw=orange,fill=yellow!40,right=1.5 of edit,thick]
      (cache) {Cache?} ;
    \draw [->,myarr] (smod.east) -- +(1,0) |- (cache) ;

    \node[sidestepbox,draw=orange,fill=orange!40,right=1.5 of run]
      (nvcc) {GPU Compiler} ;
    \draw [->,myarr] (cache) -- (nvcc) node [pos=0.4,anchor=west] {no};

    \node[objbox,draw=orange,fill=yellow!40,right=0.5 of nvcc]
      (cubin) {GPU Binary} ;
    \draw [->,myarr] (nvcc) -- (cubin) ;

    \draw [->,myarr] (cache) -| (cubin) node [pos=0.1,above=-0.1cm] {yes} ;

    \node[sidestepbox,draw=orange,fill=orange!40,below=0.5 of nvcc]
      (upload) {Upload to GPU} ;
    \draw [->,myarr] (cubin) |- (upload) ;

    \node[stepbox,draw=green,fill=green!40,below=1 of smod]
      (gpurun) {Run on GPU} ;
    \draw [->,myarr] (upload.south) -- ++(0,-0.5) -| (gpurun) ;

  \end{tikzpicture}
  \caption{Workflow of PyCUDA GPU program compilation. PyCUDA aims to maintain
    a scripting-like ``edit-run-repeat'' style of working for the user. The
    compilation and caching operations in the gray box are performed without
    user involvement.}
  \label{fig:pycuda-workflow}
\end{figure}
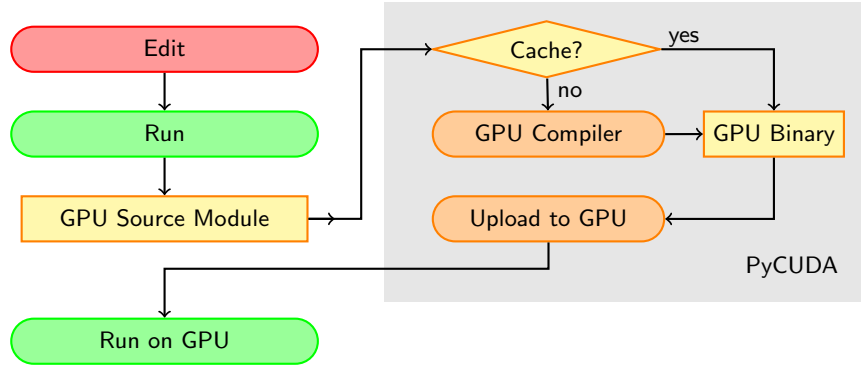

The basic shell described so far establishes the basis for more interesting,
higher-level features. PyCUDA augments the CUDA runtime system by a critical
capability: It allows the user to easily create on-GPU binaries simply by
providing C-like CUDA\footnote{For completeness, it should be mentioned that
PyCUDA also allows the just-in-time compilation of code expressed in Nvidia's
lower-level ``PTX'' abstract machine language.} source code as a simple
character string. This is what enables GPU run-time code generation.
A similar capability is available in OpenCL itself, and thereby also in
PyOpenCL.

Two factors contribute to making this process easy and transparent: First, the
user makes no contact with the underlying compiler infrastructure unless
desired.  Second, the result of the compilation process is stored in a
semi-permanent cache and reused if possible.  The cache is sensitive to changes
in the hardware and software environment and initiates recompilation when
necessary. As a result, compilation of source code and subsequent loading of
the binary code becomes nearly instantaneous and invisible to the user, and the
quick turn-around time of a scripting-based programming environment is
retained. Figure~\ref{fig:pycuda-workflow} illustrates the principle, the end
result of which is to make computations specified by C source code a library
service that is available cheaply. 

As a side observation, it is interesting to note that this way of programming
amounts to embedding one language in another, a concept that dates back to the early
days of computing \citep[e.g.][]{feldman_translator_1968} and continues to
be popular for, e.g., embedding of assembly in higher-level languages today.

Further, whenever GPU RTCG is used for automated tuning, it is desirable that
the expense of time and processing power involved in the tuning is only
incurred once per relevant code change. In most cases, the presence of a compiler
cache is already sufficient here, as compilation is usually several orders of
magnitude more time-consuming than the actual timing run of the code. However,
when that is not the case, we support the building of an application-level
cache by offering means for the easy gathering of identifying information
regarding hardware, software and their corresponding versions.

\begin{figure}[ht]
  \begin{minipage}[t]{0.48\textwidth}
    a)
\begin{lstlisting}[linerange=1-18]
import pycuda.driver as cuda
import pycuda.autoinit
import numpy

a = numpy.random.randn(4,4).astype(numpy.float32)
a_gpu = cuda.mem_alloc(a.nbytes)
cuda.memcpy_htod(a_gpu, a) # host-to-device

mod = cuda.SourceModule("""
    __global__ void multiply_by_two(float *a)
    {
      int idx = threadIdx.x + threadIdx.y*4;
      a[idx] *= 2;
    }
    """)

func = mod.get_function("multiply_by_two")
func(a_gpu, block=(4,4,1))
\end{lstlisting}

    \begin{tikzpicture}[remember picture, overlay]
      \filldraw[rounded corners, opacity=0.3, color=red] (0.3,1.95) rectangle
      +(0.95\textwidth, 1.7) node[below left=1.2cm and 0cm, text=black,opacity=1,
      font=\footnotesize\sffamily] {
        Compute Kernel};
    \end{tikzpicture}
  \end{minipage}
  \hfill
  \begin{minipage}[t]{0.48\textwidth}
    a) cont'd.
\begin{lstlisting}[linerange=20-30]
a_doubled = numpy.empty_like(a)
cuda.memcpy_dtoh(a_doubled, a_gpu) # device-to-host
print a_doubled
print a
\end{lstlisting}
    b) 
\begin{lstlisting}
import numpy
import pycuda.autoinit
import pycuda.gpuarray as gpuarray

a_gpu = gpuarray.to_gpu(
    numpy.random.randn(4,4).astype(numpy.float32))
a_doubled = (2*a_gpu).get()
print a_doubled
print a_gpu
\end{lstlisting}
  \end{minipage}

  \caption{a) An example of the use of PyCUDA, showing the use of the
  \texttt{SourceModule} facility for (static) GPU run-time code generation.
  This simple program uploads a $4\times 4$ array of single-precision
  floating point numbers, multiplies them by two on the GPU, and retrieves 
  the result.
  b) An example performing the same function as a), but using
  \texttt{GPUArray}s.
  }
  \label{fig:pycuda-demo}
\end{figure}

The combination of RTCG with services of the run-time system such as
high-precision timing and code property access already suffices to enable the
strategies laid out in Section~\ref{sec:problem-overview}.
Figure~\ref{fig:pycuda-demo}a) illustrates, by way of a sample program, how the
pieces explained so far fit together.

\subsection{Comparison to related Projects}

This may be an opportune time to compare our approach based on RTCG and
language embedding to a number of related projects seeking to ease
programming of GPU-like computer architectures.

\begin{description}

\item[CorePy] \cite{mueller_corepy_2007} was a direct inspiration to PyCUDA
(and, indirectly, PyOpenCL) in clarifying the utility of run-time code
generation. It allows the programmer to build an in-memory representation of an
assembly-language program, using machine-level instructions and various
generation utilities,
which can then be executed.

\item[jCUDA] \cite{yan_jcuda_2009} for Java is a far more literal language
mapping of the original CUDA C than PyCUDA, implementing equivalents of CUDA
C's source-to-source translation and, just like it, focusing on ahead-of-time
compilation.

\item[hiCUDA] \cite{han_hicuda_2009} could conceivably be described as
an attempt at ``OpenMP for GPU programming''. In adding an often
substantial number of \texttt{\#pragma} directives, the user enables a
source-to-source translator to rewrite an otherwise valid
single-processor C program into a GPU-executable variant. Memory
organization, data movement and execution layout are still explicit,
and much of the underlying abstraction (and complexity) is preserved
at a user-facing level, while removing some of the possibilities (such
as texture fetches). As such, we perceive hiCUDA as an alternative
syntax for CUDA kernels which relieves the user from coding explicit
data movement.

\item[BSGP] \cite{hou_bsgp_2008} is an adaptation of the bulk-synchronous
parallel (BSP) model \cite{valiant_bridging_1990} to the GPU environment.  It
implements a rather more general execution model than is supported by current
GPU hardware and makes many hardware-related choices (such as computation
layouts) for the programmer.

\item[Sh] \cite{mccool_metaprogramming_2004,mccool_data-parallel_2006} is
discussed elsewhere in this text.

\item[Brook] \cite{buck_brook_2004} is one of the earlier, very
high-level abstractions available for GPU computing. Within a C-like
language, it exposes a ``stream'' abstraction that puts rather severe
constraints on the data access pattern that a program may use,
commensurate with the capabilities of the graphics hardware it was
targeting. Brook was subsequently generalized to Brook++, which served
as an abstract programming model for AMD GPU hardware for a while.

\end{description}

\subsection{Abstractions and Convenience Functions}

One of the fundamental principles of our software is that while high-level features
are desired, their use should never obstruct access to low-level capabilities, and
their use should never obscure the underlying processes.
The purpose of this is twofold:
\begin{itemize}
  \item Uninhibited low-level access ensures that all opportunities for
    unanticipated uses of low-level facilities are retained.
  \item Whenever a high-level abstraction is used, the developer deciding to use it
    assumes a responsibility to know what the abstraction does, fix it if it breaks,
    or adapt it if is no longer suitable.
\end{itemize}
Keeping this in mind, PyCUDA and PyOpenCL include a number of abstractions, but
strive to keep them simple and ``flat''. They further strive to only include
``popular'' abstractions that are expected to be useful to a significant share
of client codes, lessening the maintenance burden on every individual user.  As
such, they exhibit a reasonably low-level interface that can and is being used
by others to create higher-level machinery (cf.
\cite{omar_cloquence_2011,catanzaro_copperhead_2011}).

\subsubsection{Numerical Arrays on the Compute Device}

Our packages provides computational linear algebra involving vectors and
multi-dimensional arrays that are designed to match the interface of the
widely-used (CPU-based) Python array package \texttt{numpy}
\cite{oliphant_numpy_2006}.  This array class, called \texttt{GPUArray} in
PyCUDA, and simply \texttt{Array} in PyOpenCL, offers a complete set of
features, including
\begin{itemize}
  \item elementwise algebraic operations such as addition, multiplication, etc.,
  \item a full set of floating-point transcendental as well as utility functions,
  \item type promotion and arbitrary combinations of data types (e.g. adding 32-bit
    integers to 32-bit floating point values results in 64-bit floating point values
    to preserve precision),
  \item reductions such as sums, maxima, and inner products, and
  \item tight integration with the \texttt{numpy} \cite{oliphant_numpy_2006}
    Python array package.
\end{itemize}
Using the array infrastructure, PyCUDA also implements GPU-based
sparse matrix-vector multiplication, as described by Garland and Bell
\cite{bell_spmv_2009}. Based on this feature, in turn, we were
able to include a fast conjugate-gradient-based \cite{hestenes_methods_1952}
linear system solver, which uses the GPU to solve large systems about ten times
faster than competing CPU implementations. Both of these facilities interact
seamlessly with the CPU-based SciPy module \cite{scipy_2001}.

\begin{figure}[ht]
  \begin{minipage}[t]{0.48\textwidth}
    a)

\begin{lstlisting}
import pycuda.autoinit
import pycuda.gpuarray as gpuarray
from pycuda.curandom import rand as curand
from pycuda.elementwise import ElementwiseKernel

x = curand((500000,))
y = curand((500000,))
z = gpuarray.empty_like(x)

lin_comb = ElementwiseKernel(
    "float a, float *x, float b, float *y, float *z",
    "z[i] = a*x[i] + b*y[i]")
lin_comb(5, x, 6, y, z)
\end{lstlisting}
  \end{minipage}
  \hfill
  \begin{minipage}[t]{0.48\textwidth}
    b)

\begin{lstlisting}
import pycuda.autoinit
import pycuda.gpuarray as gpuarray
from pycuda.curandom import rand as curand
from pycuda.elementwise import ElementwiseKernel, \
        VectorArg, ScalarArg

x = curand((500000,))
y = curand((500000,))
z = gpuarray.empty_like(x)

lin_comb = ElementwiseKernel([
    ScalarArg(x.dtype, "a"), VectorArg(x.dtype, "x"),
    ScalarArg(y.dtype, "b"), VectorArg(y.dtype, "y"),
    VectorArg(x.dtype, "z")],
    "z[i] = a*x[i] + b*y[i]")

lin_comb(5, x, 6, y, z)

\end{lstlisting}
  \end{minipage}
  \caption{
    Elementwise linear combinations
    implemented via PyCUDA's elementwise-operation code generator, accessible as
    \texttt{pycuda.elementwise.ElementwiseKernel}.
    a) shows a simple, statically typed version.
    b) shows a version that relies on type introspection to generate
    code that is appropriate for the given combination of array types.
    (The result type is defaulted to the first argument's type for
    simplicity.)
  }
  \label{fig:pycuda-elwise-demo}
\end{figure}

On top of arrays, our packages offer code generation features for custom
elementwise and reduction operations. These work by letting the user specify
only short snippets of C code for core functionality, while supplying loop
slicing and driver code automatically.  Figure~\ref{fig:pycuda-elwise-demo}a)
illustrates this for the elementwise operation case, implementing a two-vector
linear combination.  The reduction code generator is similar in spirit. We
would like to emphasize the ease with which this simple RTCG tool overcomes the
common problem of proliferation of temporary variables plaguing abstract,
operator-overloading array packages. C++ packages employing template techniques
can achieve a similar degree of efficiency through the \emph{expression
template} mechanism \cite{veldhuizen_1997}, but a robust, usable implementation
of this technique is far more complex than the simple generation of C code
involved in the RTCG solution. In general, the effort required to create RTCG
programs scales very gently with the degree of sophistication required.
Figure~\ref{fig:pycuda-elwise-demo}b) illustrates this by extending the
previous linear combination code to adapt the vector types in the generated
code
dynamically, by making use of Python's run-time type introspection. It may be
argued that these examples look pleasant only because PyCUDA contains a nice
enough pre-made user interface that suits this purpose. This is certainly true,
but the point should be seen in a different light: Only by working in a high-level
language were we able to provide this type of user interface. Since providing usable,
abstract interfaces is more straightforward in scripting environments, this
niceness becomes the rule rather than the exception.

\subsection{Code Generation}
\label{sec-code-generation-with-pycuda}
We now turn to how a user might go about creating abstractions such as
\texttt{ElementwiseKernel} herself. Since PyCUDA and PyOpenCL can natively process 
a flavor of C code, the objective is the generation of such
code. Either package makes no assumptions about the origins of the code it processes,
which allows the logic involved in the generation to be designed to match the
needs of the application. There are, however, three suggested ways of
generating code which we have found to cover a variety of needs.

\begin{figure}[ht]
  \begin{minipage}[t]{0.48\textwidth}
    a)

\begin{lstlisting}[linerange=codegen-end]
import pycuda.driver as cuda
import pycuda.autoinit
import numpy
import numpy.linalg as la
from pycuda.compiler import SourceModule

block_size = 16
thread_block_size = 32
macroblock_count = 33

a = numpy.random.randn(block_size*thread_block_size*macroblock_count)\
        .astype(numpy.float32)
b = numpy.random.randn(block_size*thread_block_size*macroblock_count)\
        .astype(numpy.float32)

a_gpu = cuda.to_device(a)
b_gpu = cuda.to_device(b)
c_gpu = cuda.mem_alloc(a.nbytes)

# codegen
from jinja2 import Template

tpl = Template("""
    __global__ void add(
            {{ type_name }} *tgt, 
            {{ type_name }} *op1, 
            {{ type_name }} *op2)
    {
      int idx = threadIdx.x + 
        {{ thread_block_size }} * {{block_size}}
        * blockIdx.x;

      {% for i in range(block_size) %}
          {% set offset = i*thread_block_size %}
          tgt[idx + {{ offset }}] = 
            op1[idx + {{ offset }}] 
            + op2[idx + {{ offset }}];
      {% endfor %}
    }""")

rendered_tpl = tpl.render(
    type_name="float", block_size=block_size,
    thread_block_size=thread_block_size)

smod = SourceModule(rendered_tpl)
# end

func = smod.get_function("add")
func(c_gpu, a_gpu, b_gpu, 
        block=(thread_block_size,1,1),
        grid=(macroblock_count,1))

c = cuda.from_device_like(c_gpu, a)

assert la.norm(c-(a+b)) == 0
\end{lstlisting}
  \end{minipage}
  \hfill
  \begin{minipage}[t]{0.48\textwidth}
    b)

\begin{lstlisting}[linerange=codegen-end]
import pycuda.driver as cuda
import pycuda.autoinit
import numpy
import numpy.linalg as la
from pycuda.compiler import SourceModule

block_size = 16
thread_block_size = 32
macroblock_count = 33
dtype = numpy.float32

a = numpy.random.randn(block_size*thread_block_size*macroblock_count)\
        .astype(dtype)
b = numpy.random.randn(block_size*thread_block_size*macroblock_count)\
        .astype(dtype)

a_gpu = cuda.to_device(a)
b_gpu = cuda.to_device(b)
c_gpu = cuda.mem_alloc(a.nbytes)

# codegen
from codepy.cgen import FunctionBody, \
        FunctionDeclaration, Typedef, POD, Value, \
        Pointer, Module, Block, Initializer, Assign
from codepy.cgen.cuda import CudaGlobal

mod = Module([
    FunctionBody(
        CudaGlobal(FunctionDeclaration(
            Value("void", "add"),
            arg_decls=[Pointer(POD(dtype, name)) 
                for name in ["tgt", "op1", "op2"]])),
        Block([
            Initializer(
                POD(numpy.int32, "idx"),
                "threadIdx.x + %d*blockIdx.x" 
                % (thread_block_size*block_size)),
            ]+[
            Assign(
                "tgt[idx+%d]" % (o*thread_block_size),
                "op1[idx+%d] + op2[idx+%d]" % (
                    o*thread_block_size, 
                    o*thread_block_size))
            for o in range(block_size)]))])

smod = SourceModule(mod)
# end

func = smod.get_function("add")
func(c_gpu, a_gpu, b_gpu, 
        block=(thread_block_size,1,1),
        grid=(macroblock_count,1))

c = cuda.from_device_like(c_gpu, a)

assert la.norm(c-(a+b)) == 0

\end{lstlisting}
  \end{minipage}
  \caption{
    Different methods of Run-Time Code Generation (RTCG) with PyCUDA.
    Example~a) generates a piece of C code from a textual template
    implementing an unrolled version of vector addition.
    (using the \texttt{Jinja2} engine \cite{ronacher_jinja_2009}
    in this instance)
    Example~b) builds a data structure approximating a C syntax tree
    for the same purpose as a).
    This tree is then converted to C code using the authors' 
    \texttt{codepy} package \cite{kloeckner_codepy_2009}.
    Full context for both examples can be found in the PyCUDA source
    tree as
    \texttt{examples/demo\_meta\_template.py.txt}
    and
    \texttt{examples/demo\_meta\_codepy.py.txt}.
  }
  \label{fig:pycuda-meta-demo}
\end{figure}

\begin{description}
  \item[Simple textual keyword replacement.] This simple technique performs the
  equivalent of search-and-replace on source code. It suffices for a
  surprisingly large range of use cases, such as the substitution of types and
  constants into source code at run time.  Its technological reach is increased
  by combining it with C preprocessor macros. Further contributing to its
  attractiveness, Python's standard library can perform keyword substitution
  without relying on external software.

  \item[Textual Templating.] For code generation applications where control
  flow and conditionals are required, but all code variants are textually
  related, the use of a so-called templating engine, commonly used for the
  generation of web pages, offers a natural escalation of the capabilities of
  keyword substitution. Many templating engines (and correspondingly,
  templating languages) exist.  Figure~\ref{fig:pycuda-meta-demo}a)
  demonstrates the use of the Jinja2 \cite{ronacher_jinja_2009} engine for the
  generation of a simple, partially unrolled vector addition code.

  \item[Syntax Tree Building.] The use of templating finds its limits
  if the codes to be generated cease to be textually related.  Then it
  may become appropriate to introduce a full representation of the
  target code in the host language in the form of a syntax tree.
  Syntax tree building allows code to be generated using all
  facilities of the host language. In particular, while templating is
  mostly ``flat'' and oriented along the lines of the output, syntax
  tree building allows the user to use, e.g., a hierarchy of functions
  to generate the desired code.

  Figure~\ref{fig:pycuda-meta-demo}b) demonstrates the use of the
  authors' CodePy \cite{kloeckner_codepy_2009} package for the
  generation of the same unrolled vector addition code as in the
  previous example. Comparing Figures \ref{fig:pycuda-meta-demo}a) and
  b) reveals that syntax tree generation code entails a
  significant departure from a form that may be easily written by
  someone familiar with the underlying C-like programming language.
  This may introduce maintenance difficulties, even though it
  is also visible that tree generation does not require much more
  generating code or a ``giant conceptual leap'' when compared to
  templating.
\end{description}

We have already emphasized various times that one of the central goals of
PyCUDA and PyOpenCL is to facilitate the construction of abstractions, the more
sophisticated of which amount to \emph{domain-specific languages}.
From a compiler construction perspective, the three strategies above
amount to using C as an intermediate representation in the building
of a compiler for such a language.


PyCUDA is available from \url{http://mathema.tician.de/software/pycuda},
PyOpenCL from \url{http://mathema.tician.de/software/pyopencl}. Both are
distributed under the liberal MIT open-source software license. Full
documentation is available online and packaged with the distribution, along
with a large body of examples and tests. The package supports all platforms on
which CUDA and/or OpenCL is available.  Both have been used in a variety of
research codes (see Section~\ref{sec:applications} for a few examples). In
addition, both packages can be used interactively from the command line as well
as from the notebook interface of the Sage exploratory computation system
\cite{stein_sage_2005}.



\section{Successful Applications}
\label{sec:applications}

\begin{figure}
  \begin{minipage}[t]{0.48\textwidth}
    a)

    \includegraphics[width=\textwidth]{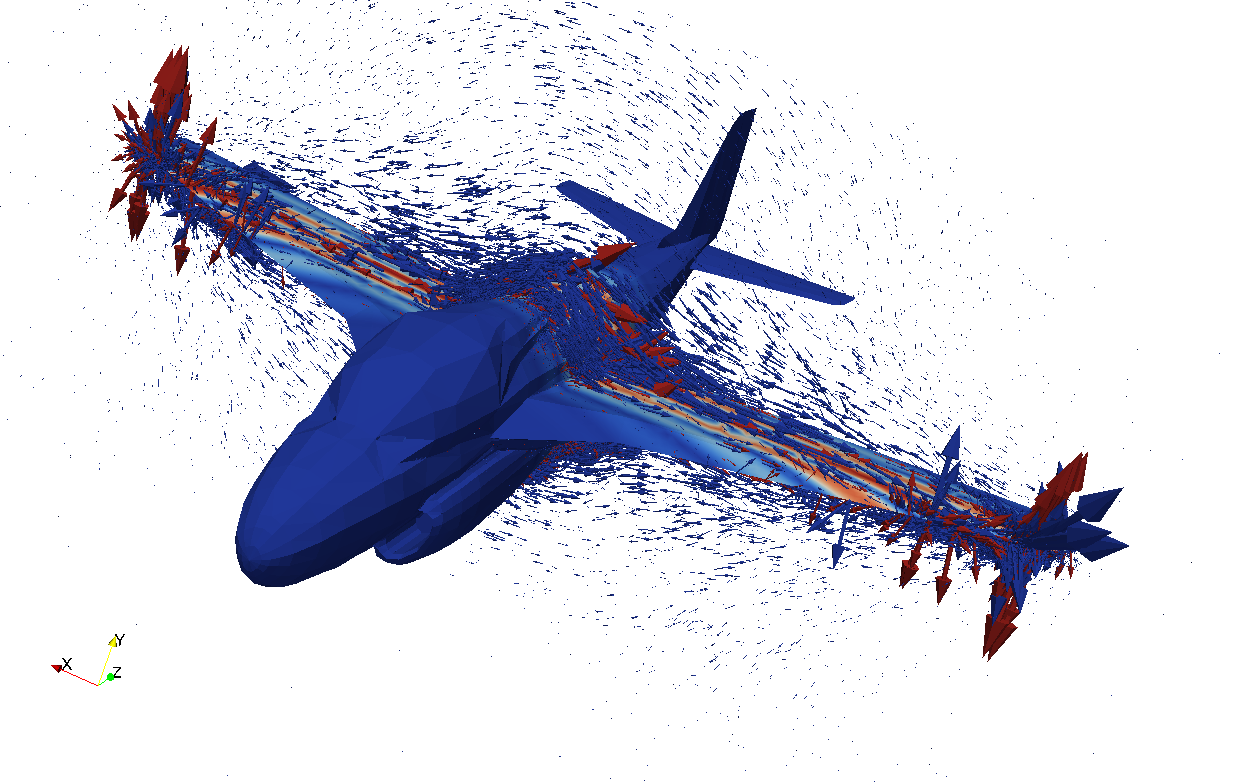}
  \end{minipage}
  \hfill
  \begin{minipage}[t]{0.48\textwidth}
    b)

    \includegraphics[width=\textwidth]{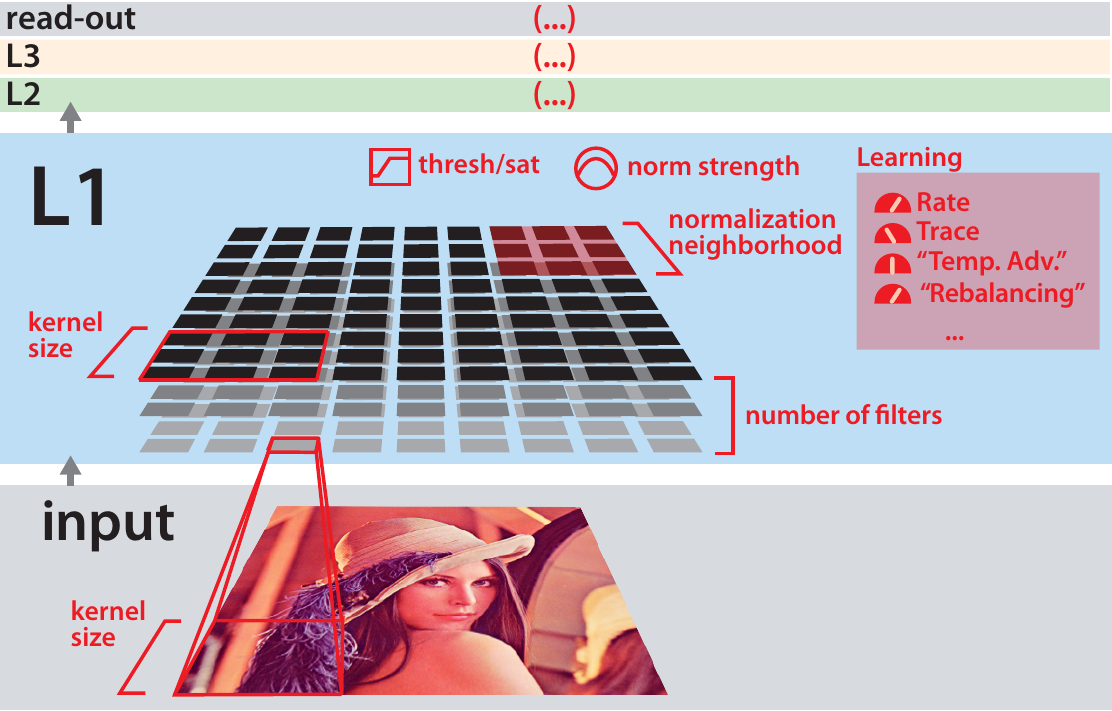}
  \end{minipage}
  \centering
  \caption{
    a)
    A sample scattering problem solved using the DG-FEM methods described
    Section~\ref{sec:applications-dg}.
    The incident plane-wave electric field is shown as pseudocolor values on
    the scatterer, while the scattered electric field is shown as arrows.  The
    computation was performed at fourth order on a mesh of 78745 elements
    using an incident-field formulation \cite{hesthaven_nodal_2002} and
    characteristic absorbing boundary conditions. It achieved and sustained
    more than 160 GFlops/s using a single Tesla C1060.
    b)
    A schematic diagram of the family of biologically-inspired computer vision
    models considered in Section~\ref{sec:applications-neuro}. The system
    architecture consists of three feedforward filtering layers, with the
    filters in each layer being applied across the previous layer. Red colored
    labels indicate a selection of configurable parameters (only a subset of the
    52 parameters are shown). Exploring this family efficiently was
    fundamentally enabled by the methods and tools described in this paper, as
    writing optimal GPU code for any model instantiation \emph{by hand} would be
    prohibitive.
  }
  \label{fig:research-samples}
\end{figure}

PyCUDA has been used successfully in a considerable number of research
projects.  We outline a few projects and their use of RTCG in detail below.
Beyond those, an up-to-date listing of successful uses of PyCUDA, PyOpenCL and
GPU run-time code generation in general can be found on the web at
\url{http://wiki.tiker.net/PyCuda/ShowCase}.

\subsection{Discontinuous Galerkin Finite Element PDE Solvers}
\label{sec:applications-dg}

Discontinuous Galerkin finite element methods (DG-FEM) for the numerical
solution of partial differential equations are popular because they are both
flexible and robust: They allow arbitrary geometries and easy control of
accuracy without compromising simulation stability. In addition to their
favorable numerical properties, DG schemes combine high arithmetic intensity,
local memory access and locally dense linear algebra. They are therefore
computationally well-suited for implementation on GPUs.  However, DG-FEM also
face significant challenges for GPU implementation, many of which were already
captured in abstract form above.

Computationally, the method operates on a long vector of degrees of
freedom, partitioned up by the element to which they belong.  Each
element may have between four and a few hundred degrees of freedom
associated with it, depending on the desired order of approximation.
The unstructured computational mesh on which the method is carried out
may then contain hundreds of thousands of individual elements. The
application of a DG operator necessitates a number of element-local
matrix-vector multiplications (by matrices of sizes between $4\times
4$ and about $300\times 300$) along with a number of non-local
inter-element operations. This operator is then applied many times
within a time-stepping loop. The major challenge is to provide code
that performs well across a broad range of orders of approximation.

A number of techniques for dealing with this problem are enumerated
and evaluated in \cite{kloeckner_2009}. We will not reiterate all the
different options here, but instead refer the interested reader to
that work for the details. Our code employs automated tuning, as described
above, to choose between these options. The first (or ``outermost'')
tuning stage concerns memory layouts.
For each given memory layout, we exploit that DG operators can be
split into of independent operations, each of which is then tuned
independently, using various choices for, e.g.  loop slicing and use
of on-chip storage. From these individual measurements, a joint score
assigned based on a target operator, and the memory layout is chosen
based on this score. While the procedure is still mainly brute-force
in nature, it employs a few heuristics to recognize poor solutions
early on. In addition, the code exploits code generation to
efficiently generalize over many different partial differential
equations and mesh dimensionalities.  In summary, we found that for
high orders of accuracy (and thus large matrices), numerous fast code
variants exist, and manual tuning is feasible (if tedious). At
lower orders, fast codes seem to be less abundant and depend on
``lucky coincidences'' that are difficult to find by hand. This
difficulty is owed in part to many of the matrices' sizes being
poorly matched to the number of SIMD lanes available.

This finding is supported by comparing with a mathematically equivalent code
which was hand-written by a colleague. This alternative code does not employ
automated tuning or RTCG.  In terms of development effort and line count, a
fair comparison is difficult to make, since the generating version is both
faster and more general. The GPU-specific parts occupy roughly 6500 lines of
code within a solver package of 33000 lines, with about 150 lines specific to
each application problem.  Each individual application problem in the
conventional version occupies perhaps 1000 lines, within a code of also 33000
lines.  When reasoning about these numbers, the reader should further consider
that the general code contains a number of alternative codes that perform the
same task, serving as alternatives in automated tuning.  With respect to
performance, the non-generating code manages to do just as well as the
generating version for large matrices (and rather very high orders), but for a
practically relevant middle range of orders (3, 4, and 5, with matrix sizes of
$20\times 20$ and $56\times 56$), the generating version fares better by a
factors of $2$, $1.6$, and $1.3$.  Further, it is encouraging to see that the
generated code achieves full memory bandwidth utilization from order 3 upwards.
For further details and performance data on the generating version, we refer
the interested reader to \cite{kloeckner_2009}.

\subsection{Computational Visual Neuroscience}
\label{sec:applications-neuro}

The study of biological vision and the creation of artificial vision systems
are naturally intertwined as they represent simultaneous efforts to forward and
reverse engineer systems with similar goals. However, while neuroscience has
provided inspiration for some of the ``broad-stroke'' properties of natural
visual systems, much is still unknown.  As a result, we are often left
exploring a staggeringly large hypothesis \emph{space} of models, rather than
evaluating any one model \emph{per se}.  To pave a way forward, we have
recently developed a high-throughput approach \cite{pinto_plos_2009} to more
expansively explore the possible range of brain-inspired models (which consist,
roughly speaking, of a cascade of multiple layers of regular parallel
computations such as linear filtering operations and static nonlinearities, see
Figure~\ref{fig:research-samples}b), including models of larger, more realistic
scale by leveraging recent advances in commodity stream processing hardware. In
analogy to high-throughput screening approaches in molecular biology, we
generate and train thousands of potential model instantiations, and ``screen''
their visual representations using an object recognition task. From these
candidate models, the most promising are selected for further analysis. We have
shown that this approach can yield significant, reproducible gains in
performance across an array of object and face recognition tasks, consistently
outperforming a variety of state-of-the-art purpose-built vision systems from
the literature, and that it can offer insight into which computational ideas
are most important for achieving this performance \cite{pinto_plos_2009,
pinto_bionetics_2010, pinto_fg_2011}.

The brain itself is a highly parallel statistical supercomputer, and thus
algorithms inspired by its function are well suited to the computational
advantages offered by GPUs. This power naturally comes at the cost of increased
complexity for the developer (Section~\ref{sec:problem-overview}), and the need
to explore a wide range of different kinds of biologically-inspired models
poses a serious challenge for optimization.  Optimization is often an exercise
in specialization: an algorithm is carefully matched to set of
hardware/software resources, exploiting as much regularity in the underlying
problem and inputs as possible. However, for our problem, we must build
algorithms that can tolerate widely varying input domains since a major drive
of our work is to find model parameter sets that can provide high levels of
performance on many different tasks (like the human visual system).  Ideally,
we would like to have the best possible implementation in each context, without
having to undertake a massive effort in hand-tuning.  

In the last five years, our group has experienced three different
``paradigms'': programming GPUs with graphics
primitives\footnote{\url{http://www.opengl.org},
\url{http://developer.nvidia.com/page/cg_main.html},
\url{http://www.gpgpu.org}} in 2006, programming the PlayStation 3 using
low-level Cell Broadband Engine
intrinsics\footnote{\url{http://www.ibm.com/developerworks/power/cell/index.html}}
in 2007 and programming GPUs with compute primitives like CUDA\footnote{
\url{http://developer.nvidia.com/object/gpucomputing.html}} in 2008.  To
overcome the challenge of optimizing for each architecture, we applied RTCG to
auto-tune the core operations by ``instrumentalizing'' low-level code and
manipulating it with a Python template engine
(Figure~\ref{fig:pycuda-meta-demo}b).  We implemented common optimization
strategies (e.g. loop unrolling \cite{kennedy_opt_compilers_2001}, pre-fetching
and software pipelining \cite{lam_systolic_1989}, alleviation of register
pressure using spilling \cite{wang_register_1994}, communication and
computation load distribution, etc.) and achieved comfortable speed-ups with a
simple auto-tuning method (i.e. coarse grid search, see below). In the future,
we plan to investigate the use of machine learning techniques for auto-tuning,
an approach recently undertaken by IBM's Milepost GCC \cite{ibm_milepost_2008}.

Using RTCG with Python, we were able to combine the flexibility and ease-of-use
of a high-level language for ``outer loop'' control and auto-tuning, with the
raw performance of highly optimized ``close-to-the-metal'' GPU or Cell code to
achieve hundred-fold speedups over conventional MATLAB/MEX CPU implementations
(the standard in the fields of computational neuroscience and computer vision;
see \cite{pinto_plos_2009}'s Figure~1, S1 and Text S2 for more details about 
these comparisons and for an in-depth discussion about the impact of our RTCG
approach on programmability and productivity). We argue that the combination of
these qualities enables a new kind of exploration of ideas in biological and
computational sciences, where scale is matched with the fluid ability to
experiment new ideas.

Below, we illustrate the power of RTCG toolkits like PyCUDA to perform simple
empirical auto-tuning using a key computational bottleneck in
our algorithms: 3D filter-bank convolution (which mimics one part of
the processing that is thought to be done by biological neurons). Note that
this \emph{regular} operation is arguably straightforward to optimize when all
of the relevant constraints of the hardware/software stack are known in advance
and the inputs' dimensions are fixed, which is not the case in our particular
research application. 

For the purposes of demonstration, we chose a large set of simple optimization
configurations (i.e. unique combinations of loop unrolling depth, register
spilling, block/grid dimensions, thread work size, shared memory padding, etc.)
and auto-tuned for four different inputs, which roughly bracket the range of
possible input shapes and sizes that the are encountered in our experiments.
The results of RTCG auto-tuning, on various NVIDIA GPUs spanning multiple
generations of graphics hardware, multiple end-user markets (gaming versus
professional), and a wide range of variation in hardware-level resources
available, are shown in Table~\ref{tab:fbconv3_autotuning}.  Large performance
gains are observed for the GPU programs generated and tuned
\emph{automatically} as compared to the ``default'' GPU program, which was
laboriously \emph{hand-coded} and \emph{hand-tuned} to allow (optimal) correct
execution of all input ranges on all GPUs -- without running up against
hardware limitations.

\begin{table}[ht]
\input{fbconv3_perf_table.tex}

\caption{
Performance of RTCG auto-tuning on 3D filter-bank convolutions using PyCUDA.
}
\label{tab:fbconv3_autotuning}
\end{table}

Interestingly, further analyses of the data (see \cite{pinto_gcg_2011} for
details) show that a different peak-performing optimization configuration was
chosen for each input size, and different for distinct hardware platforms.
Given the many demands on system resources that trade-off against each other, a
different ``sweet-spot'' implementation exists for different incoming inputs
and for different constellations of hardware resources. We also observe large
differences in performance (in some cases over 100\%) when a custom hardware
RTCG auto-tuned GPU program is used, as compared to when an optimal GPU program
for a different platform is used.  Such performance differences are
particularly important when development is done on a different machine (e.g. a
laptop) than where the code will be run in production mode.  Similarly, for
applications that are widely deployed on a variety of user hardware, optimal
performance can be achieved by either optimizing \emph{in situ} or shipping
with a database of optimization configurations for different platforms.
Without meta-programming, hand-tuning for each of the many hardware
configuration in existence and for many different input configurations would be
a tedious and error-prone process.  By contrast, RTCG, in combination with a
trivial auto-tuning scheme, allow optimal implementations to be chosen on any
platform and input size (see \cite{pinto_gcg_2011} for more information).

As the scale of available computational power continues to expand, and more
RTCG tools like PyCUDA emerge, we believe that this approach has the potential
to greatly accelerate progress in both artificial vision and our understanding
of the computational underpinning of biological vision.  \emph{(N. Pinto, J.J.
DiCarlo, David D. Cox)}

\subsection{Copperhead}
\label{sec:applications-copperhead}
RTCG also serves as the foundation of higher level programming tools
and environments, such as Copperhead.  Copperhead
\cite{catanzaro_copperhead_2011} is a data parallel language embedded
in Python, with the goal to improve programmer productivity while
still providing excellent computational performance.  Using
Copperhead, programmers express computation in terms of composition of
data-parallel primitives, such as \texttt{map}, \texttt{reduce},
\texttt{gather} and \texttt{scatter}.  Copperhead is implemented as
a standard Python library that uses RTCG to map compositions of data
parallel primitives onto GPU hardware.  An embedded source-to-source
compiler creates CUDA code which implements the desired computation,
which is then compiled and executed on the GPU.  PyCUDA manages
lazy data transfers to and from the GPU, as well as all GPU memory
resources, thanks to its efficient memory pool facility which avoids
extraneous calls to \texttt{cudaMalloc} and \texttt{cudaFree} when
repeatedly reallocating data of similar shapes.  

Although systems like Copperhead could conceivably be implemented
using static compilation approaches, RTCG brings three crucial
benefits.  Firstly, it obviates the standard static compilation
workflow, which impedes productivity, especially for programmers who
ordinarily use scripting languages such as Python.  The details of
invoking static compilers are abstracted away cleanly using RTCG, and
there is no need to write compilation scripts in order to build
programs using RTCG.  Although this is a minor theoretical advantage,
in practice we find this to be quite useful, since many programmers
are accustomed to working with scripting languages and find
traditional compilation flows daunting.  This impedes the adoption of
new architectures such as GPUs which are not supported directly by
today's scripting languages, and is overcome by RTCG.  Secondly, as
attested elsewhere in this paper, RTCG naturally enables autotuning
and design space exploration.  The Copperhead compiler must make
mapping and scheduling decisions during the process of lowering a set
of nested data-parallel primitives into a parallel program running on
a particular processor.  With RTCG, an autotuner can instruct the
compiler to generate multiple variants of the same code, in order to
ensure efficiency.  Thirdly, with RTCG, the compiler can examine
selected properties of the inputs to a Copperhead function, and
specialize the resulting code for those inputs.  This must be done
with caution, in order to avoid extraneous invocation of the static
compiler, which is costly, but can be very useful; for example, when
the compiler attempts to allocate data to onchip memories on the GPU,
knowledge of data shapes and types is very helpful. Summarizing, RTCG
has proven an essential tool for Copperhead's infrastructure.

\begin{figure}[ht]
\begin{minipage}{0.38\textwidth}
\begin{lstlisting}[linerange=1-15]
from copperhead import *
import numpy as np

@cu
def axpy(a, x, y):
    def triad(xi, yi):
        return a * xi + yi
    return map(triad, x, y)

a = np.float32(np.random.randn())
n = 1000000
x = np.random.randn(n, 1).astype(np.float32)
y = np.random.randn(n, 1).astype(np.float32)

z = axpy(a, x, y)
\end{lstlisting}
      \caption{A simple example Copperhead program, which performs
        scaled vector addition.}
      \label{fig:copperhead_demo}
\end{minipage}
\hfill
\begin{minipage}{0.6\textwidth}
    \includegraphics[width=.98\textwidth]{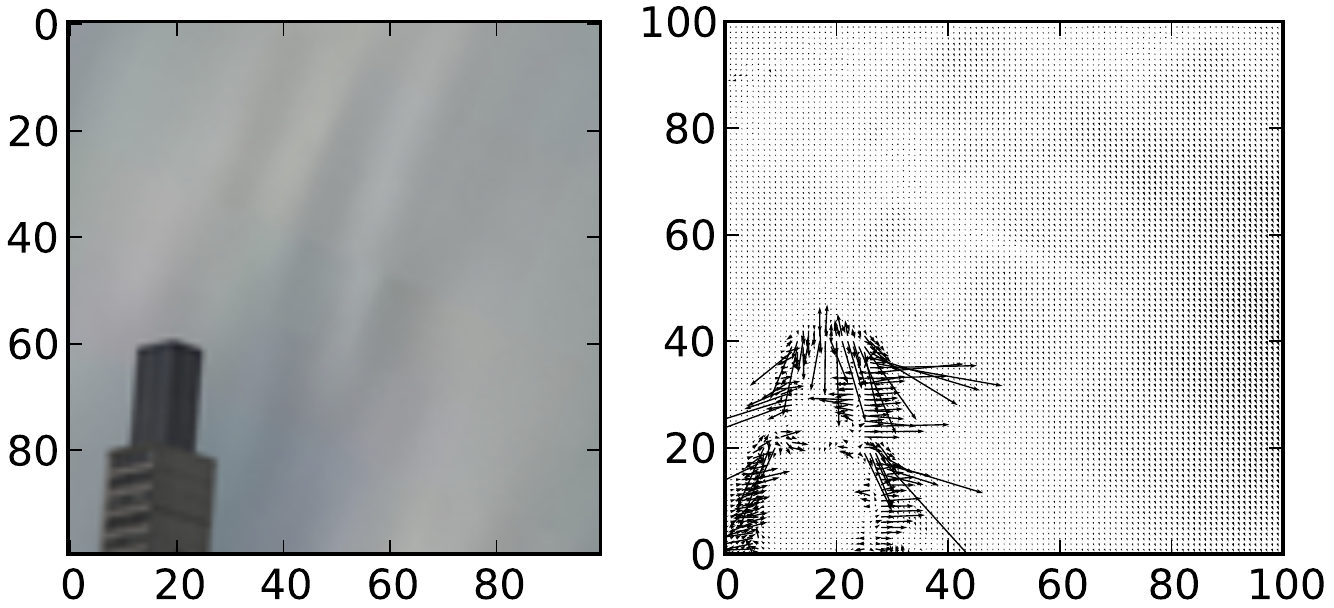}
    \caption{Step in Non-linear Optical Flow solver computed and
      visualized using Copperhead}
    \label{fig:copperhead_plot}
 \end{minipage}
\end{figure}
Figure \ref{fig:copperhead_demo} shows a simple Copperhead program
which performs scaled vector addition. The abstraction ability of
today's scripting languages such as Python allows for the compilation
and data movement required to run programs on the GPU to be hidden in
the Copperhead library. By carefully mapping nested data parallel
computations onto the GPU, the Copperhead compiler is able to achieve
fairly high performance - within 45-100\% of handcoded CUDA programs.
Since Copperhead is embedded in Python, it can interoperate with
standard Python libraries for numeric and scientific computing, such
as {\texttt numpy}, {\texttt scipy}, and {\texttt matplotlib}.  This
makes Copperhead a productive environment for implementing entire
programs, not just their computationally intensive kernels. For
example, the graph in figure \ref{fig:copperhead_plot} was created
during execution of a Copperhead program using {\texttt matplotlib}.

\begin{table}[ht]
  \begin{minipage}[t]{0.48\textwidth}
  \centering
  \begin{tabular}{|l|c|c|}
   \hline
   Example & CUDA & Copperhead \\
   & Perf. & Perf.  \\
   & (GFLOP/s) & (GFLOP/s) \\ \hline
   CSR Scalar SpMV & 1.8  & 1.8 \\ \hline
   CSR Vector SpMV & 12.0 & 5.5 \\ \hline
   ELL SpMV & 13.5 & 10.5 \\ \hline
   PCG Solver & 34 & 24.5 \\ \hline
   SVM Solver & 71 & 36 \\ \hline
  \end{tabular}
  \caption{Copperhead performance versus Hand-written CUDA performance}
  \label{tab:copperhead_performance}
  \end{minipage}
  \hfill
  \begin{minipage}[t]{0.48\textwidth}
  \centering
  \begin{tabular}{|l|c|c|}
  \hline
   Example & CUDA & Copperhead \\ 
   & LOC & LOC \\ \hline
   CSR Scalar SpMV & 16 & 6 \\ \hline
   CSR Vector SpMV & 39 & 6 \\ \hline
   ELL SpMV & 22 & 4 \\ \hline
   PCG Solver & 172 & 79 \\ \hline
   SVM Solver & 429 & 111 \\ \hline
   \end{tabular}
   \caption{Standardized Lines of Code comparison between Copperhead
     and CUDA}
   \label{tab:copperhead_productivity}
   \end{minipage}

\end{table}

Performance results of some sample Copperhead programs are shown in
Table~\ref{tab:copperhead_performance}.  As mentioned earlier, we achieve between 45-100\%
of hand-coded CUDA performance (see \cite{catanzaro_copperhead_2011}
for more details.)

In Table~\ref{tab:copperhead_productivity}, we show the size of some
example Copperhead programs, as well as the equivalent handwritten
CUDA C++ code.  Productivity cannot be directly measured using
standard-lines-of-code counting, however we believe line counts are
correlated to programmer productivity.  Accordingly, we believe that
the fact that Copperhead programs require, on average, approximately 4
times fewer lines of code than CUDA C++ programs, may indicate that it
is a more productive way to program parallel processors like
GPUs. Copperhead is enabled by RTCG, and we believe there will be many
other frameworks which utilize RTCG to provide programmers with highly
productive, efficient ways to take advantage of modern parallel
processors like GPUs.  Copperhead is available at
\url{http://code.google.com/p/copperhead} under the Apache 2.0
license.  \emph{(B. Catanzaro, M. Garland, K. Keutzer and Y. Lee)}

\subsection{Estimating the Entropy of Natural Scenes}
\label{sec:applications-natscenes}
Characterizing the statistics of natural scenes is an important area of vision
research. The entropy of images provides a measure of the information content
available to the visual system and as such quantifies the demands placed on
neural information processing mechanisms.  From an
applications perspective, entropy is the theoretical limit of
compression--the lower bound on {\em any} compression scheme.  Recently,
\cite{chandler_estimates_2007} used an entropy estimation algorithm to
binlessly estimate the entropy of small patches of natural images from the
distribution of nearest-neighbor distances. 

The main computational bottleneck involves finding, for each $8\times8$ image
patch in a target set, its Euclidean distance nearest neighbor in a neighbors
set.  Due to the high-dimensional nature of the data and the requirement for
finding exact and not just approximate nearest neighbors, we are limited to
using an exhaustive approach of calculating the distance of each target to each
of the neighbors, and taking the smallest of these. Furthermore, the method
requires for the neighbors set to double in size with each iteration, so we
need to calculate the distance to every neighbor patch in an exponentially
growing set.  We mitigate this limitation by parallelizing the brute force
nearest neighbor search on the GPU. 

PyCUDA gave me an efficient manner with explore and experiment with different
approaches.  As a GPU novice, when I started CUDA programming in C, the
overhead of recompiling and keeping track of different versions of the code
greatly slowed down experimentation with using different kinds of memory
(global, shared, texture, register), access patterns, and computational
strategies. Exploration is essential for learning the performance
landscape, but I lacked the means for doing such exploration efficiently.  RTCG
let me concentrate on writing compute kernels, instead of keeping track of
makefiles, with code generation and compilation conveniently abstracted away.

\begin{table}
\centering
	\begin{tabular}{ | r | r | r | r || c | c |}
\hline
neighbors     & PyCUDA      &PyCUDA    &C\hspace{1.5em} &speedup &speedup\\
      &  (8800GTX) &(GTX 295)&     (gcc -O)    & (8800 GTX)& (GTX 295)\\
\hline
4096  &0.144 s     &  0.089 s&   3.76 s&25.95    &42.25\\
16384 &0.521 s     &  0.299 s&  15.03 s& 28.83   &50.27 \\
65536 &2.047 s     &  1.146 s&  60.16 s& 29.39   &52.50 \\
262144 & 8.036  s      &  4.508 s& 242.13 s&  30.13      & 53.79 \\
1048576& 32.093 s      & 17.989 s&    969.00  s &  30.19      &53.87\\
\hline

\end{tabular}
\caption{ The computational time and speedup comparison for finding the nearest
      neighbor of 4096 target patches from among different numbers of neighbors.  
      Each patch is $64$ dimensional ($8\times8$).
The C implementation ran on one core of a 2.4 GHz Intel Core2 Q6600.
}
\label{tab:NN-speedup}
\end{table}

One Nvidia 8800GTX card performs 30 times faster than a compiler optimized C version
and the speedup increased to being 53 times faster on one of the two GPUs in a
GTX295 without any code changes (see Table~\ref{tab:NN-speedup} for details).
Reproducing the results in \cite{chandler_estimates_2007} for $2^{18}$
neighbors takes 3 hours using our CPU implementation, but the same computation
takes just 3.2 or 6 {\em minutes} depending on the GPU used.  The speedups
obtained enable us to perform more extensive entropy and fractal
dimensionality analyses on the entire database of about 4000 thousand natural
images\cite{van_hateren_independent_1998} in a single day, whereas this would
take more than a month using the CPU, which is why only a few dozen images were
used in the previous work.  Additionally, because our implementation uses
PyCUDA, we can easily optimize the parameters of the implementation for newer
cards, and extend the parallelism to multiple cards.  Such computational
capabilities will enable us to analyze and compare previously unimaginable
large classes of images in a reasonable amount of time. \emph{(P. Ivanov)}

\subsection{Filtered Backprojection for Radar Imaging}
\label{sec:radar}

Tomographic systems as diverse as x-ray CT and synthetic aperture radar use a
sensor that projects a three-dimensional function (such as x-ray absorption and
electromagnetic reflectivity of a scene, respectively) onto one-dimensional
range profiles. Reconstruction of the original function from a collection of
these line projections can be accomplished by the filtered backprojection
algorithm, which for the simpler two-dimensional problem is:
$$
    I[n_x,\, n_y] = \sum_{m=1}^M D[m,\, r] \cdot e^{j u r},
$$
where a two-dimensional image $I$ is indexed by integer pixels $n_x$ and $n_y$
for some imaging grid, and where each of the $M$ range profiles is a row in the
complex-valued data matrix $D$. This data matrix is indexed by the profile
number $m$, and a factional range bin $r \equiv r(n_x,\, n_y,\, p_x[m],\,
p_y[m],\, p_w[m])$, an arithmetic function of the image pixel in question, the
two-dimensional location of the sensor at each projection $(p_x[m],\, p_y[m])$,
and the distance between the sensor and the scene center at each projection
$p_w[m]$. Each image pixel must query each projection---a row in the data
matrix---for its scalar contribution, and apply a phase shift proportional to a
sensor-specific $u$ (potentially variable across projections, i.e., $u\equiv
u[m]$) and $r$ before summing. Here, $j=\sqrt{-1}$.

This algorithm can be seen as $M$ interpolation steps, and may be implemented
as a triple loop indexing $n_x$ and $n_y$, the output image axes, and $m$, the
projections, giving $\mathcal{O}(MN^2)$ complexity for an $N\times N$ image.
The interpolation at each pixel-and-projection pair requires a different
fractional index, and as precomputing them all poses a significant memory
burden, they must be calculated within the inner-most loop. Fetching irregular
data samples from memory for weighting is the major obstacle that
implementations need to overcome.

Thanks to the large number of execution units and linear interpolation hardware
available on modern GPUs, a mildly-optimized CUDA implementation performs SAR
backprojection over 50 times faster on a C1060 Tesla than a single-threaded
implementation on a modern CPU \cite{Fasih-Hartley-Radar-2010}.  The range
profile data matrix is loaded into texture memory to take advantage of both
caching and linear interpolation. The three or more time series required by the
algorithm (the position and range of the sensor and the potentially variable
phase offset) may be loaded into texture memory or uncached global memory. The
output image is partitioned to blocks, with each thread responsible for a
single pixel, implementing the ``inner'' loop over range profiles. After
processing all range profiles, the image is copied, in block-sized chunks, back
to GPU global memory or pinned host memory (with little difference between
these two). 

The irregular unpatterned nature of the algorithm's memory accesses, the
indeterminism of texture caching, and register pressure make tuning a CUDA
implementation essential. Orientation of the two-dimensional data matrix in
texture memory ($D[m,\, r]$ versus $D[r,\, m]$), location of the per-projection
sensor time series, location of the output (global memory to host memory versus
pinned host memory), and of course block sizes are all non-kernel
implementation factors that affect throughput. 

The algorithm was implemented for use from Python via PyCUDA as well as from
MATLAB through the MEX interface, which enables MATLAB to interact with
compiled C programs, in our case, an all-C CUDA backprojection implementation
In both these cases, preliminary data processing was done on the host. However,
facilities to load the data and recover the result were provided by PyCUDA,
whereas these steps required manual coding for the MEX version. Special array
conversion code was also required for the MEX version, whereas this was
provided for by PyCUDA and the \texttt{numpy} Python package. 

The use of newer C or C++ packages that abstract away some of this bookkeeping
would have simplified the MEX code. However, the numerous edit-compile-run
cycles needed to tune the MEX implementation are an inherent part of this
toolchain, whereas the PyCUDA version allowed for the more rapid interactive
command-line experimentation expected from a Python environment.

Also note that a cleaner and simpler kernel is obtained by the use of
pre-compiled constants for the numerous imaging and sensor parameters, rather
than passing these in as function arguments. Programmatic modification of the
source code to update such constants is much more natural in Python, per
Section~\ref{sec-code-generation-with-pycuda}, than in MATLAB, where the
transformed source code would be written to the disk and explicit compile
commands invoked in the shell.

For reasons outlined above, the C source code required by the PyCUDA
implementation was shorter than the C source code for the CUDA-enabled MEX
implementation, as well as the benchmark single-threaded CPU MEX
implementation.
\begin{itemize}
    \itemsep 1pt \parskip 0pt
    \item CPU MEX implementation: 570 lines of C source code, excluding header lines
    \item CUDA MEX implementation: 420 lines
    \item PyCUDA implementation: 115 lines
\end{itemize}
Although runtimes between the PyCUDA and the CUDA MEX implementations were not
systematically measured for the same imaging scenarios, users did not report a
performance discrepancy between the two implementations. Runtime comparisons
between the separate CUDA implementations and the CPU implementation, as well
as the implementation tuning needed for two imaging scenarios, were given in
\cite{Fasih-Hartley-Radar-2010}.  \emph{(A. Fasih)}



\section{Conclusions}
\label{sec:conclusions}

We have described the powerful consequences of the confluence
of two events in high-performance computing: First, the emergence of
general-purpose programmable GPUs as a viable mass market product has made
performance jumps of an order of magnitude or more a reality for a number of
important applications. Second, the maturing of open-source scripting languages
and their software ecosystems has enabled similar jumps in productivity for
creators of scientific software. It is straightforward to see that a hybrid
model combining GPUs and scripting offers numerous advantages over more
traditional models of software creation.

The main message of this paper is that through the natural addition of GPU
run-time code generation to this mixture, one automatically
combines the strengths and compensates for the weaknesses of each of the
technologies involved, leading to a compelling way of constructing
high-performance computational software.

To make GPU RTCG accessible, we have built, documented, and published PyCUDA
and PyOpenCL, two toolkits that allow the easy application of the principles
described here. We have described the facilities available and demonstrated
their use.  We will continue to extend and maintain both PyCUDA and PyOpenCL.

Based on these toolkits, we will explore the construction of tools that allow
researchers to focus on their target areas, while leaving the detailed work
involved in accomplishing basic computational tasks to the machine.  One effort
that is currently underway will use empirical optimization to try and find
well-performing kernels for a certain set of basic array operations, such as
those involved in dense numerical linear algebra or certain PDE solvers.
Further, it should not be forgotten that PyCUDA and PyOpenCL were born out of
the need of actual applications, as Section~\ref{sec:applications} illustrated.
As the research in these application areas progresses, we fully expect that
more advanced needs will drive the implementation of even better tools.

In summary, we believe that the flexibility of run-time generated code
provides a crucial tool in unlocking the performance capabilities of advanced
hardware to a broader mass of developers, and we look forward to the
opportunities and challenges that future hardware generations will bring.

\subsection*{Acknowledgments}

The authors would like to thank the PyCUDA and PyOpenCL communities, without
whom both projects would not be where they are today.

AK would first and foremost like to thank his doctoral advisor
Jan~Hesthaven at Brown~University for providing constant encouragement
and a stimulating environment.  Jan~Hesthaven and Xueyu~Zhu at Brown
read the manuscript and contributed many improvements.  AK would also
like to thank Tim~Warburton at Rice~University and Michael~Garland at
Nvidia for the valuable advice they provided along the way. Last, but
not least, he would like to acknowledge Nvidia Corporation, who, upon
completion of this work, provided Brown University with a generous
hardware donation for further research. AK's research was partially
funded by AFOSR under contract number FA9550-07-1-0422, through the
AFOSR/NSSEFF Program Award FA9550-10-1-0180 and also under contract
DEFG0288ER25053 by the Department of Energy.  The opinions expressed
are the views of the authors.  They do not necessarily reflect the
official position of the funding agencies.

NP would like to thank James J. DiCarlo, David D. Cox, Steven
G. Johnson and Hanspeter Pfister for helpful discussions--as well as David
Luebke, Joe Stam, John Roberts and Nvidia for their support in both his research
and teaching.

AF would like to thank \"{U}mit~\c{C}ataly\"{u}rek and Tim~Hartley for their
collaboration as well as Randy~Moses, Emre~Ertin, and Josh~Ash, all of The Ohio
State University, for technical discussion and financial support.  Finally, AF
gladly acknowledges the support of the Ohio Supercomputer Center.



\bibliography{gpu-rtcg}
\bibliographystyle{abbrvnat}
\end{document}

%% file: fbconv3_perf_table.tex
{\small
\begin{tabular}[c]{|c|c|c|c|c|c|}

\multicolumn{6}{c}{} \\
\hline
\textbf{\emph{GPU}} & 
\textbf{\emph{Input}} & 
\textbf{\emph{Filter-bank}} & 
\textbf{\emph{Default}} & 
\textbf{\emph{RTCG auto-tuned}} & 
\textbf{\emph{Boost}} \\ 
\textbf{\emph{SDK}} & 
 & 
 &
\emph{GFLOP/s} & 
\emph{GFLOP/s} & 
\\
\hline\hline
\multirow{4}{0.14\linewidth}{%
8600GT CUDA2.3
}
&
256x256x8
&
64x9x9x8
&
5.493 $\pm$ 0.019
&
33.881 $\pm$ 0.068
&
516.8 \%
\\
\cline{2-2}
\cline{3-3}
\cline{4-4}
\cline{5-5}
\cline{6-6}
&
512x512x4
&
32x13x13x4
&
11.619 $\pm$ 0.007
&
33.456 $\pm$ 0.045
&
187.9 \%
\\
\cline{2-2}
\cline{3-3}
\cline{4-4}
\cline{5-5}
\cline{6-6}
&
1024x1024x8
&
16x5x5x8
&
19.056 $\pm$ 0.017
&
33.109 $\pm$ 0.632
&
73.7 \%
\\
\cline{2-2}
\cline{3-3}
\cline{4-4}
\cline{5-5}
\cline{6-6}
&
2048x2048x4
&
4x8x8x4
&
23.824 $\pm$ 0.055
&
38.867 $\pm$ 0.118
&
63.1 \%
\\
\cline{2-2}
\cline{3-3}
\cline{4-4}
\cline{5-5}
\cline{6-6}
\hline\hline
\multirow{4}{0.14\linewidth}{%
9400M CUDA3.1
}
&
256x256x8
&
64x9x9x8
&
2.177 $\pm$ 0.013
&
15.796 $\pm$ 0.049
&
625.6 \%
\\
\cline{2-2}
\cline{3-3}
\cline{4-4}
\cline{5-5}
\cline{6-6}
&
512x512x4
&
32x13x13x4
&
5.562 $\pm$ 0.001
&
15.331 $\pm$ 0.004
&
175.6 \%
\\
\cline{2-2}
\cline{3-3}
\cline{4-4}
\cline{5-5}
\cline{6-6}
&
1024x1024x8
&
16x5x5x8
&
2.309 $\pm$ 0.022
&
4.571 $\pm$ 0.015
&
98.0 \%
\\
\cline{2-2}
\cline{3-3}
\cline{4-4}
\cline{5-5}
\cline{6-6}
\hline\hline
\multirow{4}{0.14\linewidth}{%
C1060 CUDA2.3
}
&
256x256x8
&
64x9x9x8
&
104.188 $\pm$ 0.051
&
168.083 $\pm$ 0.372
&
61.3 \%
\\
\cline{2-2}
\cline{3-3}
\cline{4-4}
\cline{5-5}
\cline{6-6}
&
512x512x4
&
32x13x13x4
&
125.739 $\pm$ 0.109
&
234.053 $\pm$ 0.266
&
86.1 \%
\\
\cline{2-2}
\cline{3-3}
\cline{4-4}
\cline{5-5}
\cline{6-6}
&
1024x1024x8
&
16x5x5x8
&
144.279 $\pm$ 0.764
&
243.697 $\pm$ 0.346
&
68.9 \%
\\
\cline{2-2}
\cline{3-3}
\cline{4-4}
\cline{5-5}
\cline{6-6}
&
2048x2048x4
&
4x8x8x4
&
180.060 $\pm$ 0.018
&
322.328 $\pm$ 0.348
&
79.0 \%
\\
\cline{2-2}
\cline{3-3}
\cline{4-4}
\cline{5-5}
\cline{6-6}
\hline\hline
\multirow{4}{0.14\linewidth}{%
GTX295 CUDA2.3
}
&
256x256x8
&
64x9x9x8
&
126.563 $\pm$ 0.590
&
262.848 $\pm$ 0.176
&
107.7 \%
\\
\cline{2-2}
\cline{3-3}
\cline{4-4}
\cline{5-5}
\cline{6-6}
&
512x512x4
&
32x13x13x4
&
172.701 $\pm$ 0.014
&
317.108 $\pm$ 0.056
&
83.6 \%
\\
\cline{2-2}
\cline{3-3}
\cline{4-4}
\cline{5-5}
\cline{6-6}
&
1024x1024x8
&
16x5x5x8
&
104.972 $\pm$ 0.011
&
168.298 $\pm$ 0.174
&
60.3 \%
\\
\cline{2-2}
\cline{3-3}
\cline{4-4}
\cline{5-5}
\cline{6-6}
&
2048x2048x4
&
4x8x8x4
&
120.693 $\pm$ 0.020
&
226.534 $\pm$ 0.195
&
87.7 \%
\\
\cline{2-2}
\cline{3-3}
\cline{4-4}
\cline{5-5}
\cline{6-6}
\hline\hline
\multirow{4}{0.14\linewidth}{%
GTX480 CUDA3.2
}
&
256x256x8
&
64x9x9x8
&
523.316 $\pm$ 8.677
&
623.759 $\pm$ 13.754
&
19.2 \%
\\
\cline{2-2}
\cline{3-3}
\cline{4-4}
\cline{5-5}
\cline{6-6}
&
512x512x4
&
32x13x13x4
&
872.353 $\pm$ 12.375
&
1002.976 $\pm$ 7.685
&
15.0 \%
\\
\cline{2-2}
\cline{3-3}
\cline{4-4}
\cline{5-5}
\cline{6-6}
&
1024x1024x8
&
16x5x5x8
&
634.110 $\pm$ 0.411
&
667.912 $\pm$ 0.364
&
5.3 \%
\\
\cline{2-2}
\cline{3-3}
\cline{4-4}
\cline{5-5}
\cline{6-6}
&
2048x2048x4
&
4x8x8x4
&
387.524 $\pm$ 0.176
&
811.660 $\pm$ 0.212
&
109.4 \%
\\
\cline{2-2}
\cline{3-3}
\cline{4-4}
\cline{5-5}
\cline{6-6}
\hline
\end{tabular}
}
\\